# Analysis of Insect-Plant Interactions Affected by Mining Operations, A Graph Mining Approach

Ali Bayat
School of Industrial and Systems Engineering
Tarbiat Modares University
Tehran, Iran
a_bayat@modares.ac.ir

Mohammad Heydari
School of Industrial and Systems Engineering
Tarbiat Modares University
Tehran, Iran
m_heydari@modares.ac.ir

Amir Albadvi*
School of Industrial and Systems Engineering
Tarbiat Modares University
Tehran, Iran
albadvi@modares.ac.ir

*Abstract*— The decline in ecological connections signifies the potential extinction of species, which can be attributed to disruptions and alterations. The decrease in interconnections among species reflects their susceptibility to changes. For example, certain insects and plants that rely on exclusive interactions with a limited number of species, or even a specific species, face the risk of extinction if they lose these crucial connections. Currently, mining activities pose significant harm to natural ecosystems, resulting in various negative environmental impacts. In this study, we utilized network science techniques to analyze the ecosystem in a graph-based structure, aiming to conserve the ecosystem affected by mining operations in the northern region of Scotland. The research encompasses identifying the most vital members of the network, establishing criteria for identifying communities within the network, comparing, and evaluating them, using models to predict secondary extinctions that occur when a species is removed from the network, and assessing the extent of network damage. The novelty of our study is utilization of network science approaches to investigate the biological data related to interactions between insects and plants.

## I. Introduction

This research aims to investigate the interplay between insects and plants from a network perspective, identify crucial individuals that play significant roles in this interaction, and explore how communication is influenced by external factors such as the presence of mines and mining activities. Mining activities have a long history and are driven by the persistent human demand for minerals and energy. However, they are often associated with severe environmental problems, leading to the destruction of plant and animal life, loss of biodiversity, and various short-term and long-term economic and social consequences. Consequently, there is now a greater emphasis on undertaking restoration efforts in mining-affected lands compared to the past, which involves planting vegetation and rejuvenating the ecosystem. Nonetheless, existing methods for evaluating these effects have limitations, and environmental restoration activities can sometimes inadvertently cause further transformations and significant changes in the environment. By closely examining environmental changes and the alteration of communication networks between plants and insects, it becomes possible to restore the environment more effectively in areas impacted by mining activities. Merely planting vegetation in mining-affected environments is insufficient for comprehensive and proper restoration, as the extinction of an insect species in an area can potentially led to the extinction of a plant species. Restoration efforts should encompass a broader network of insects and their host plants. The substantial budgets allocated for revitalizing mining areas, if not utilized to foster sustainable changes, can perpetuate ongoing transformations over time, failing to facilitate true environmental restoration and potentially exacerbating the situation further. Hence, a critical question that deserves attention in the realm of mining activities is how to effectively transform the biological network and the animal and plant species impacted by mining to facilitate accurate environmental restoration.

## II. Background

In the Estercuel locality in northeastern Spain (Iberian Peninsula), Santos et al.[1] examined plant-insect interactions from the late Early Cretaceous (latest Albian). These interactions involve two types of land-dwelling angiosperms and Klitzschophyllites, a basal eudicot species considered one of the earliest potential members of aquatic Ranunculales discovered thus far. Sender et al.[2] study examines specimens of the extinct seed fern Sagenopteris from the Lower Cretaceous in Alcaine village, Teruel Province, northeastern Spain. The research focuses on categorizing arthropod-induced plant damage types (DTs) for 75 specimens of this plant species. These leaflets are found in deposits associated with coastal fluvial and lacustrine environments connected to an Albian delta-estuary system. Santos et al.[3] presented new evidence of plant-insect interactions discovered in the Late Pennsylvanian period in the northern region of the Iberian Peninsula (León, Spain). Through their investigation of 216 fossil plant specimens, they have identified nine distinct Damage Types (DTs) indicative of these interactions. The interactions involve four Functional Feeding Groups (FFGs), including margin feeding (DT12 and DT13), hole feeding (DT09), galling (DT33, DT80, and DT116), and oviposition (DT67, DT100, and DT102), observed on Pteridophytes, Pteridospermatophytes, and Coniferophytes. Tamura et al.[4] made predictions and assessed plant-insect interactions using limited datasets and investigated the interaction between the crop plant rice (Oryza sativa) and two species of mirid bugs (Stenotus rubrovittatus and Trigonotylus caelestialium) by utilizing observational data. By employing adaptive network models, Maia et al[5] examined the ability of plant-pollinator and plant-herbivore networks to withstand species loss. Their focus was on understanding the impact of key differences in natural history between these systems, such as the demographic outcomes of interactions and the extent of generalization, which influence the potential for rewiring. They investigated how these factors contribute to the resilience of ecological networks to extinctions. Additionally, they developed a standardized measure to assess the influence of network structure on resilience by simulating extinctions in theoretical networks with controlled structures. Balmaki [6] study introduced methodological approaches for assessing parameters of insect-pollen networks using pollen samples obtained from insect specimens housed in museums. These methods offer valuable insights into the spatial and temporal dynamics of pollen-insect interactions. They serve as a complementary tool alongside other techniques used in the study of pollination, such as observing pollinator networks and conducting flower enclosure experiments. The article includes illustrative data from butterfly pollen networks spanning the last century in the Great Basin Desert and Sierra Nevada Mountains in the United States. Lewinsohn [7] focused on the long-term research conducted on the interactions between Asteraceae plants and flowerhead-feeding insects in Brazil. their research treated host species as independent entities to assess local and turnover components of herbivore diversity and expanded to investigate entire interactive communities of host plants and their associated herbivores across different localities and regions, leading to the exploration of new research avenues. Feng[8] presented findings of plant-insect interactions within the flora, obtained through a comprehensive analysis of insect-induced damage on plant specimens. In total, we identified 8 distinct types of damage caused by insects, categorized into 5 functional feeding groups, across 11 plant species. Meineke [9] utilized machine learning techniques to examine historical insect-plant interactions recorded on digitized herbarium specimens. Martins [10] conducted a global literature review on ecological methods and indicators used for the recovery and monitoring of ecosystems following mining activities. Guan [11] conducted a bibliometric analysis on the evolution of the field of ecological restoration over the past thirty years. Also Feng [12] reviewed effects of surface coal mining and land reclamation on soil properties. Joll [13] employed a network analysis methodology to measure the frequency of insect interactions with the flowers of plant species found in the Great Lakes dune ecosystem. This study by Moudrý [14] compares the effectiveness of using unmanned aerial vehicle (UAV) imagery and airborne Light Detection and Ranging (LiDAR) data during both leaf-off and leaf-on seasons for evaluating the terrain and vegetation structure of a post-mining site. The goal is to assess the potential of these technologies in monitoring hazards and gauging the success of restoration efforts. The findings provide insights into the prospects of using UAV imagery and LiDAR for effective monitoring and restoration evaluation of post-mining landscapes. Cagua [15] proposed the concept of structural controllability, which provides a means to measure the degree to which network topology can be utilized to achieve a desired state. Their approach enables the quantification of a species' control capacity, indicating its relative significance within the network. Additionally, it helps identify the specific species that are crucial in this context due to their highest potential for control. To demonstrate its application, they examine ten plant-pollinator communities, comparing those that have not been invaded with those that have. Their findings indicate that the controllability of a community is determined by the asymmetrical nature of its mutual dependencies, rather than its invasion status. The decline in ecological connections signifies the potential extinction of species. One factor contributing to this decline is disturbances and alterations in the environment. The number of connections a species possesses demonstrates its sensitivity to changes. For instance, certain insects and plants that rely on exclusive interactions with a limited number of species, or even a specific species, face a significant risk of extinction if these connections are severed. Evaluating the robustness of the network and characteristics such as the extent of species interconnections can help assess the network's vulnerability to the loss of a particular species. Additionally, it highlights the potential for cascade effects, wherein the loss of one connection leads to the loss of other connections within the network [16]. Complex networks can offer quicker predictions for certain disturbances, such as changes in vegetation [17]. In the last two decades, people's attention to preserving the quality of the environment has gradually increased and this phenomenon has affected the design and activities of mining.

### III. RESEARCH METHODOLOGY

A very living organism relies on its environment and other organisms for essential biological processes and resources. Group cohesion is vital for survival, and the study of these interdependencies falls within the realm of ecology. Different species in nature must possess a proper understanding of their environment and other organisms to fulfill their fundamental needs and adapt accordingly. Therefore, comprehending the environment, the interactions among organisms, and their impacts on each other holds great practical and scientific significance. This research aims to utilize network analysis tools to explore and understand these relationships. The primary objective of this study is to examine how changes in land use affect the network structure of insects and plants. Each case in this research will focus on specific goals among them:

- Gaining knowledge about and exploring the interconnectedness of insects and plants
- Investigating the alterations that occur in the insect-plant network because of mining activities.
- Offering a fresh perspective on the preservation and restoration of ecosystems that undergo changes because of mining operations.

The data collection for this research is conducted through observational methods, employing simple random sampling and quantitative techniques. Inferential analysis will be employed to analyze the collected data. The implementation of this research has been carried out using the R programming language, while certain aspects have also been implemented using PHP programming to address the novelty of the covered topics. Also, The Gephi software has been employed for the visualization of graphs and networks. The research methodology will proceed as follows: Firstly, data on the relationships between insects and plants will be collected. Subsequently, a network representation of the data will be constructed, and a comprehensive examination and analysis will be conducted from a network perspective. Additionally, considering the specific characteristics of the available data, two distinct sites will be selected, one near the studied mine and the other situated at a considerable distance. These sites will be treated as separate networks, allowing for a comparative analysis of network factors to gain deeper insights into the relationships and identify crucial members from various viewpoints. The primary objectives of this study were as follows: 1) To assess the effects of mining activities on the interactions between insects and plants, 2) To identify ecologically significant species within the ecosystem and explore the possibility of categorizing them, 3) To determine which species, if removed from the ecosystem, would have the most detrimental impact on its current state, 4) To prioritize the preservation of specific species during and after mining operations, aiming to mitigate the ecological damage caused by mining activities, and 5) Detecting concealed patterns within hidden communities in an insect-plant network using various famous community detection algorithms.

## IV. DATA

In graph analytics research, having a comprehensive dataset is crucial for drawing meaningful and insightful conclusions. Regrettably, research on the insects-plant interactions affected by mining operations using graph mining approaches, is limited, making it imperative to focus on this area and gather additional datasets to address this knowledge gap. The ecosystem encompasses the assemblage of living organisms within a given environment, along with all the elements and components of that environment. In essence, the ecosystem can be defined by the environment itself and its living organisms. Among the vital components of the ecosystem, the relationship between insects and plants, known as pollination, holds great significance. Pollination is a crucial process in both natural and agricultural ecosystems, playing a pivotal role in food production and sustaining human livelihood. Human dependency on natural ecosystems and agricultural systems underscores the importance of understanding and analyzing the bilateral relationship between insects and plants. In this research, an endeavor is made to employ graph mining methods to analyze this intricate relationship. To accomplish this, the available data pertaining to this relationship will be prepared and organized to establish a network representation.

Table 1 - Dataset Fields

| Site | latitude | longitude | Treatment |
|---|---|---|---|
| Date | Wind | Temperature | Order |
| Family | Insect species | Interactions | Flower Species |

Prior to initiating the analysis, it is essential to establish precise definitions for the nodes and edges, or relationships, within the graph. In the specific problem discussed, there are two distinct groups: plants, consisting of ninety species, and insects. Communication is observed when insects interact with plants, whether for pollination or feeding. As these connections originate from insect-related nodes and extend towards plant-related nodes, a bipartite network naturally emerges, with plants forming one side and insects occupying the other side.

## V. GRAPH MINING

In this section, graph mining and network visualization methods will be employed to enhance our understanding of the research objectives.

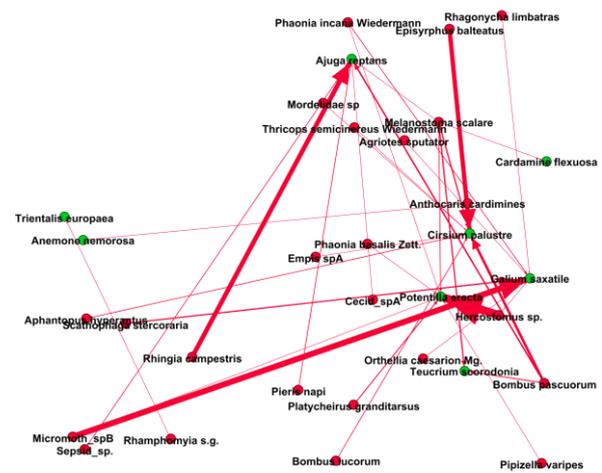

Figure 1 - Visualization of Insect-plant Relationships of Backkhill Site (Green nodes are plant species and red nodes are insect species)

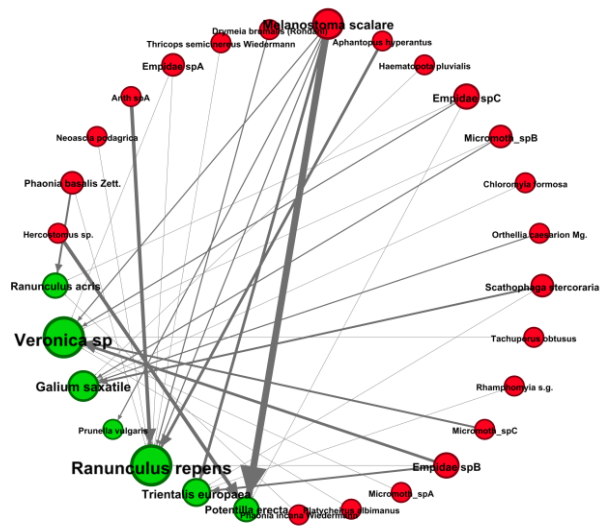

Figure 2 - Visualization of Insect-plant Relationships of Dalhaikie Site (Green nodes are plant species and red nodes are insect species)

The network under consideration is inherently a bipartite network that encompasses ninety distinct types of plants and insects. Hence, the relationships governing this network can be effectively represented through the utilization of bipartite graph.

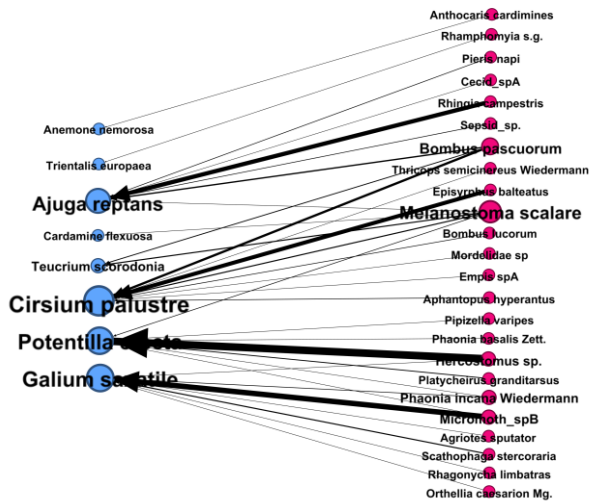

Figure 3 – Bipartite Network Visualization for Backhill site

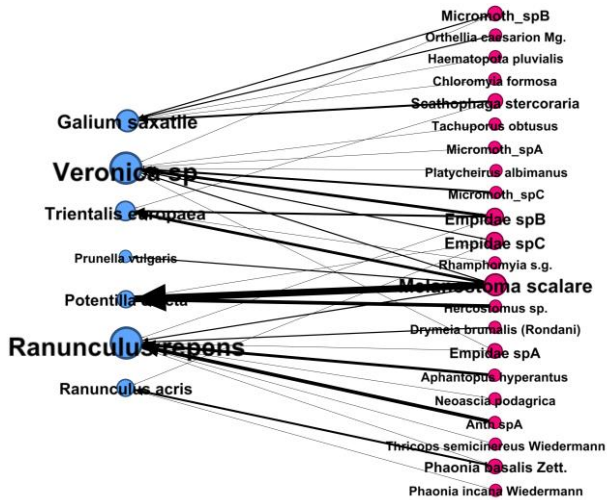

Figure 4 - Bipartite Network Visualization for Dalhaikie site

One approach to identifying significant members of the network is by utilizing centrality measures, which are discussed in this section. Initially, the simplest centrality measure, known as degree centrality, will be computed. Degree centrality solely considers the number of direct neighbors each species has and can be calculated without modifying the bipartite graph of insects and plants. Once the degree centrality values are obtained, they are sorted in descending order, resulting in a ranking of species based on their number of connections. Species with high ranks are deemed important members of the ecosystem, as they interact with numerous other network members, and their presence in the network is integral to the communication of other species. However, the significance of species with very low degrees raises questions. The species at the lower end of the sorted list represent insects and plants that possess a limited number of connections. These species are particularly vulnerable and face the risk of endangerment if they lose their pollinators or host plants. Therefore, it is imperative to prioritize the conservation of species that interact with those with very low degree centrality. The accompanying figures depict the communication network of insects and plants in two sites, Backhilland Dalhaike, respectively. Remarkably, 68.75% and 55.17% of the connections in these networks are represented by species with only one connection, indicating a high degree of privacy.

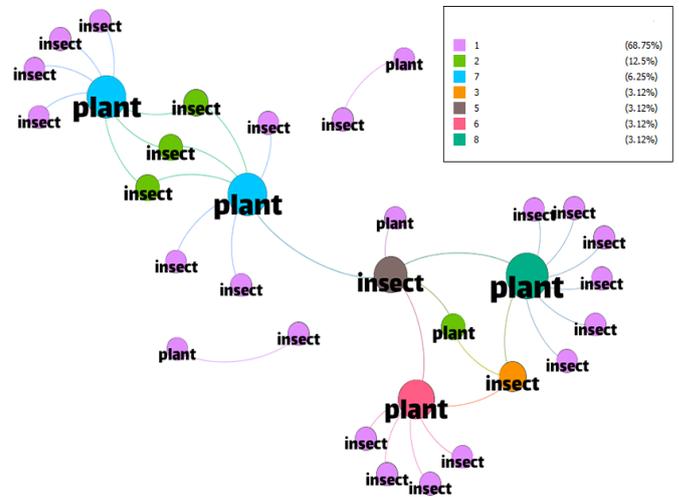

Figure 5 - Insect network of Backhill site plants by degree, (bigger size of nodes means greater degree centrality)

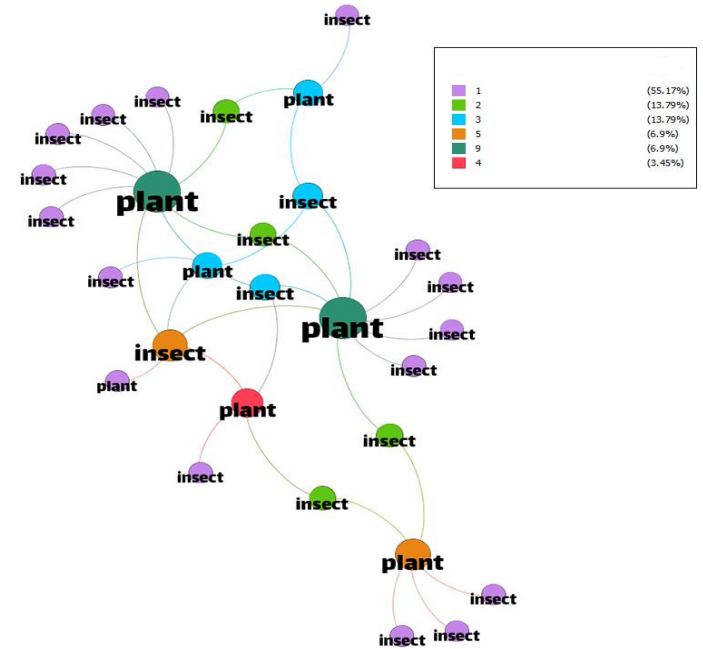

Figure 6 - Insect network of Dalhaikie site plants visualization by degree

The tables below present the top ten species that play a crucial role in establishing network connections within the insect-plant network in the Backhill and Dalhaike sites, respectively.

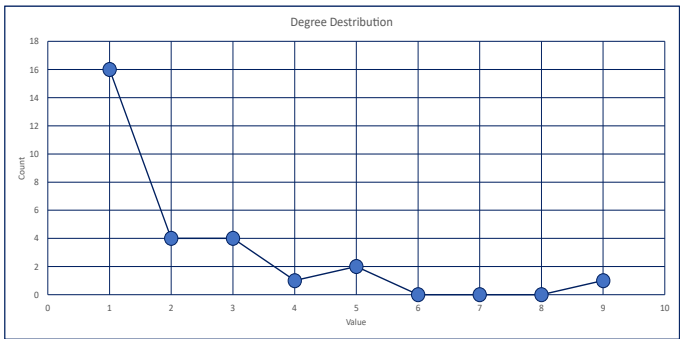

Figure 7 - Degree distributions of insect-plant network, Dalhaike site

Table 2 - Ten species with the highest degree centrality and ten vulnerable species of the Backhill site on the left and right, respectively

| Species | Deg Cent | Species | Deg Cent |
|---|---|---|---|
| Cirsium palustre | 8 | Episyrphus balteatus | 1 |
| Potentilla erecta | 7 | Rhingia campestris | 1 |
| Galium saxatile | 7 | Sepsid_sp. | 1 |
| Ajuga reptans | 6 | Empis spA | 1 |
| Melanostoma scalare | 5 | Thricops semicinereus Wiedermann | 1 |
| Bombus pascuorum | 3 | Aphantopus hyperantus | 1 |
| Hercostomus sp. | 2 | Bombus lucorum | 1 |
| Micromoth_spB | 2 | Platycheirus granditarsus | 1 |
| Phaonia incana Wiedermann | 2 | Mordelidae sp | 1 |
| Teucrium scorodonia | 2 | Phaonia basalis Zett. | 1 |

Table 3 - Ten species with the highest degree centrality and ten vulnerable species of the Dalhaike site on the left and right, respectively

| Species | Deg Cent | Species | Deg Cent |
|---|---|---|---|
| Ranunculus repens | 9 | Hercostomus sp. | 1 |
| Veronica sp | 9 | Neoascia podagrica | 1 |
| Melanostoma scalare | 5 | Anth spA | 1 |
| Galium saxatile | 5 | Thricops semicinereus Wiedermann | 1 |
| Trientalis europaea | 4 | Drymeia brumalis (Rondani) | 1 |
| Empidae spB | 3 | Aphantopus hyperantus | 1 |
| Empidae spC | 3 | Haematopota pluvialis | 1 |
| Potentilla erecta | 3 | Chloromyia Formosa | 1 |
| Ranunculus acris | 3 | Orthellia caesarion Mg. | 1 |
| Phaonia basalis Zett. | 2 | Tachuporus obtusus | 1 |

As previously mentioned, the bipartite graph representing the relationships between insects and plants can be transformed into two separate unipartite graphs. By doing so, the insect-insect relationship can be defined based on the shared plants between two insects. Conversely, the plant-plant relationship can be determined by considering the common insects involved in the pollination process of plants.

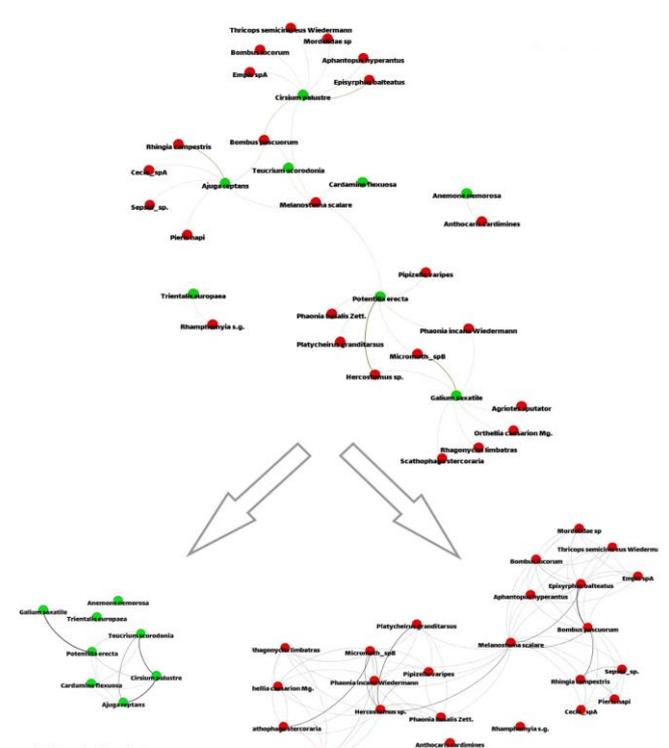

Figure 8 - Construction of insect-insect and plant-plant network from the original insect-plant network by projection method.

After converting the bipartite networks of both sites into unipartite networks for insect-insect and plant-plant relationships, the special vector centrality and closeness centrality measures were computed. In the insect-insect network, the relationship is established based on the shared plants between two insects. Consequently, an insect with a high special vector centrality is surrounded by other insects that visit numerous plants. This indicates that the insects feeding on this insect visit a diverse range of plants, suggesting a more generalist feeding behavior. However, this insect may also serve as a crucial pollinator for specific plants. Plants that exhibit a strong association with this insect, as indicated by a high special vector value, are likely to be more dependent on this specific insect for pollination. Conversely, this insect and the associated plants may be more vulnerable to damage if the insect population is disrupted, as the feeding insects responsible for pollination exhibit a greater diversity in their feeding habits. The same principle applies to the plant-plant relationship. The figures below depict the insect-insect and plant-plant unipartite graphs, showcasing the values of special vector centrality for each site.

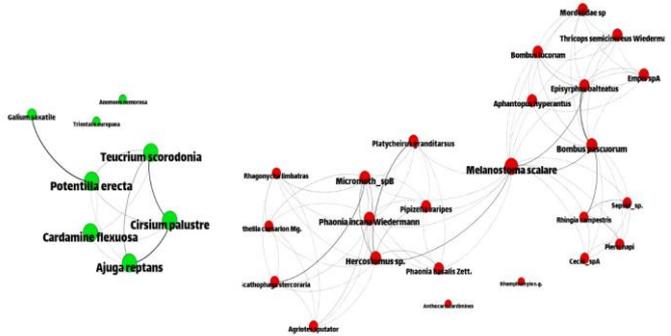

Figure 9 - Insect-insect and plant-plant one-part graphs visualization based on the value of the Eigenvector centrality of each site.

The tables provided below present the top five insects and top five plants, determined through the calculation of eigenvector centrality, in the insect-insect and plant-plant unipartite graphs of the Backhill and Dalhaike sites, respectively. These species hold the highest eigenvector centrality values, indicating their significant influence within the respective networks.

Table 4 - Backhill site eigenvector values for the five nodes with the highest eigenvalue in the insect-insect and plant-plant networks are on the left and right, respectively.

| Species | Eigenvector Centrality | Species | Eigenvector Centrality |
|---|---|---|---|
| Melanostoma scalare | 1 | Potentilla erecta | 1 |
| Bombus pascuorum | 0.646536 | Cirsium palustre | 0.950212 |
| Hercostomus sp. | 0.622216 | Ajuga reptans | 0.950212 |
| Micromoth_spB | 0.622216 | Teucrium scorodonia | 0.950212 |
| Phaonia incana Wiedermann | 0.622216 | Cardamine flexuosa | 0.950212 |

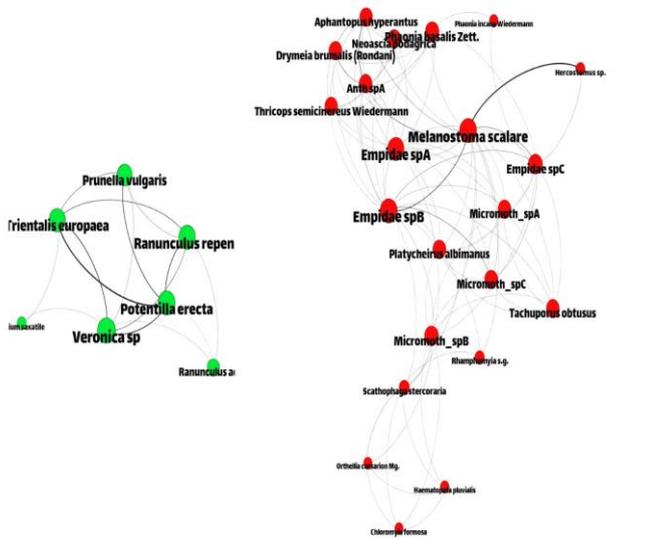

Figure 10 - Dalhaike single-mode networks, the size of the nodes represents the magnitude of the value of the eigenvector of the node.

Table 5 - Dalhaike site eigenvector values for the five nodes with the highest eigenvalue in the insect-insect and plant-plant networks are on the left and right.

| Species | Eigenvector Centrality | Species | Eigenvector Centrality |
|---|---|---|---|
| Melanostoma scalare | 1 | Veronica sp | 1 |
| Empidae spB | 0.984336 | Potentilla erecta | 0.925863 |
| Empidae spA | 0.930921 | Ranunculus repens | 0.925863 |
| Empidae spC | 0.695282 | Trientalis europaea | 0.888878 |
| Micromoth_spB | 0.680335 | Prunella vulgaris | 0.950212 |

Also, in the figures below, the insect-insect and plant-plant single-section graphs of two sites, Backhill and Dalhaike, are displayed based on the value of closeness centrality in each site.

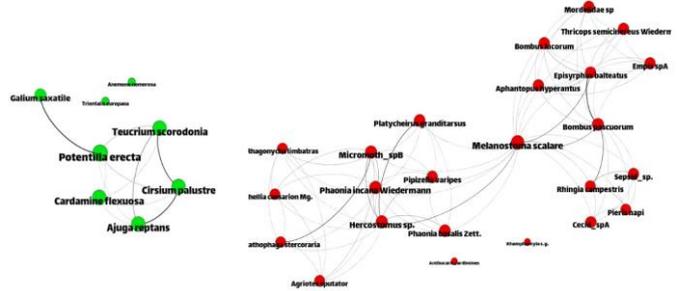

Figure 11 - Single-mode networks of the backheel site are the size of the nodes representing the magnitude of the value of the closeness centrality of the node.

Table 6 - The values of the closeness centrality of the Backhill site for the five nodes with the highest closeness centrality in the insect-insect and plant-plant networks are on the left and right.

| Species | Closeness Centrality | Species | Closeness Centrality |
|---|---|---|---|
| Melanostoma scalare | 0.84 | Potentilla erecta | 1 |
| Hercostomus sp. | 0.65625 | Cirsium palustre | 0.83333 |
| Micromoth_spB | 0.65625 | Ajuga reptans | 0.83333 |
| Phaonia incana Wiedermann | 0.65625 | Teucrium scorodonia | 0.83333 |
| Bombus pascuorum | 0.6 | Cardamine flexuosa | 0.83333 |

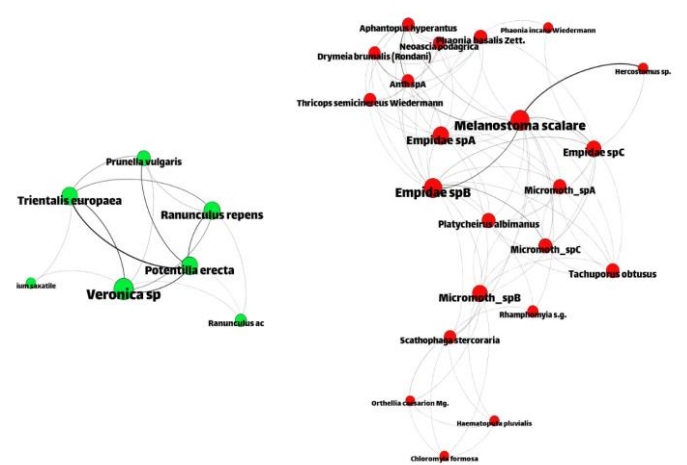

Figure 12 - Dalhaike single-mode networks, the size of the nodes represents the magnitude of the closeness centrality values of the node.

Table 7 - The closeness centrality values of Dalhaike site for the five nodes with the highest closeness centrality in the insect-insect and plant-plant networks are on the left and right.

| Species | Closeness Centrality | Species | Closeness Centrality |
|---|---|---|---|
| Melanostoma scalare | 0.84 | Veronica sp | 1 |
| Empidae spB | 0.8096 | Potentilla erecta | 0.8571 |
| Empidae spA | 0.75 | Ranunculus repens | 0.8571 |
| Micromoth_spB | 0.7 | Trientalis europaea | 0.8571 |
| Empidae spC | 0.6774 | Prunella vulgaris | 0.75 |

By utilizing the value calculation algorithm, species can be ranked based on their association with private species in the network. The algorithm emphasizes that species linked to a larger number of private species possess higher value, signifying their significance in preserving such private species. The results of applying the value calculation algorithm to two specific sites, Backhilland Dalhaike, are presented in the tables below. These tables exhibit the top ten species with the highest calculated values for each site. Additionally, the graphs illustrate a comparative depiction of the values across different types of networks for the Backhilland Dalhaike sites.

Table 8 - Ten species with the highest value obtained from the value calculation algorithm on the Backhillsite.

| species | type | node worth |
|---|---|---|
| Bombus pascuorum | insect | 0.02399 |
| Melanostoma scalare | insect | 0.05862 |
| Rhamphomyia s.g. | insect | 0.03030 |
| Anthocaris cardimines | insect | 0.03030 |
| Cirsium palustre | plant | 0.19798 |
| Ajuga reptans | plant | 0.13737 |
| Potentilla erecta | plant | 0.14242 |
| Galium saxatile | plant | 0.16667 |
| Trientalis europaea | plant | 0.03030 |
| Anemone nemorosa | plant | 0.03030 |

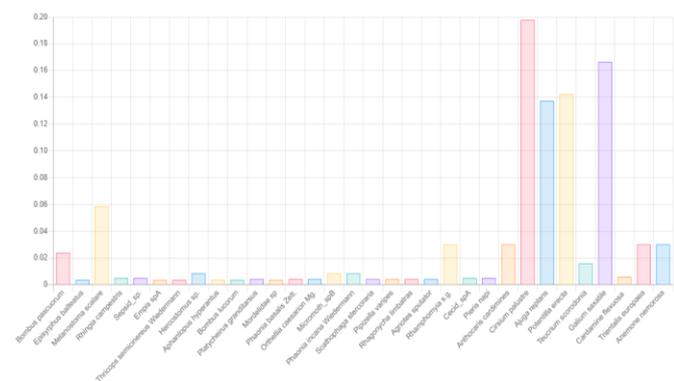

Figure 13 - The amount of value calculated in the value calculation algorithm for all types of Backhill sites.

Table 9 - Ten species with the highest value obtained from the value calculation algorithm on the Dalhaike site.

| species | type | node worth |
|---|---|---|
| Melanostoma scalare | insect | 0.05310 |
| Empidae spB | insect | 0.01389 |
| Empidae spC | insect | 0.02288 |
| Scathophaga stercoraria | insect | 0.01324 |
| Potentilla erecta | plant | 0.04510 |
| Trientalis europaea | plant | 0.05980 |
| Ranunculus repens | plant | 0.19216 |
| Galium saxatile | plant | 0.11765 |
| Veronica sp | plant | 0.17255 |
| Ranunculus acris | plant | 0.05392 |

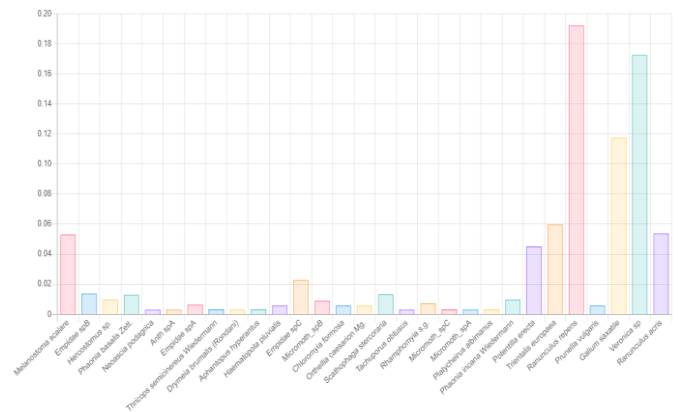

Figure 14 - The amount of value calculated in the value calculation algorithm for all types of Backhillsites.

## VI. COMMUNITY DETECTION

To evaluate the grouping of insects and plants at the Backhillsite, five different Community Detection methods were utilized. The effectiveness of each method was assessed by calculating the contract value, which represents the quality of the resulting groupings. The table presents the number of communities identified and the corresponding contract scores for each of the five methods. By comparing these values, the most suitable method can be determined.

Table 10 - Modularity score and the number of communities discovered in different methods for the Backhillsite.

| C.D Methods | Louvain | Fast Greedy | Label Propagate | ALO | FEV |
|---|---|---|---|---|---|
| **Modularity** | 0.62 | 0.62 | 0.61 | 0.62 | 0.62 |
| **Num. of C** | 5 | 5 | 6 | 5 | 5 |

In the following, the communities obtained in each method for the Backhill site are displayed.

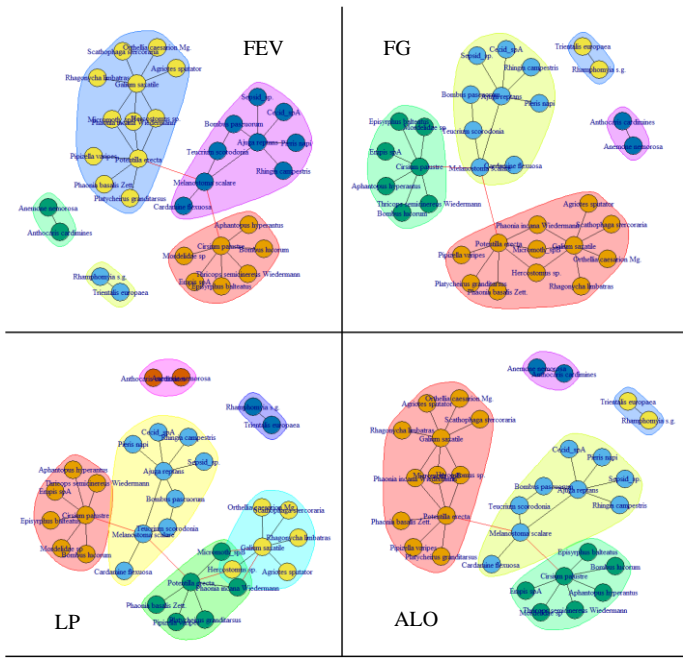
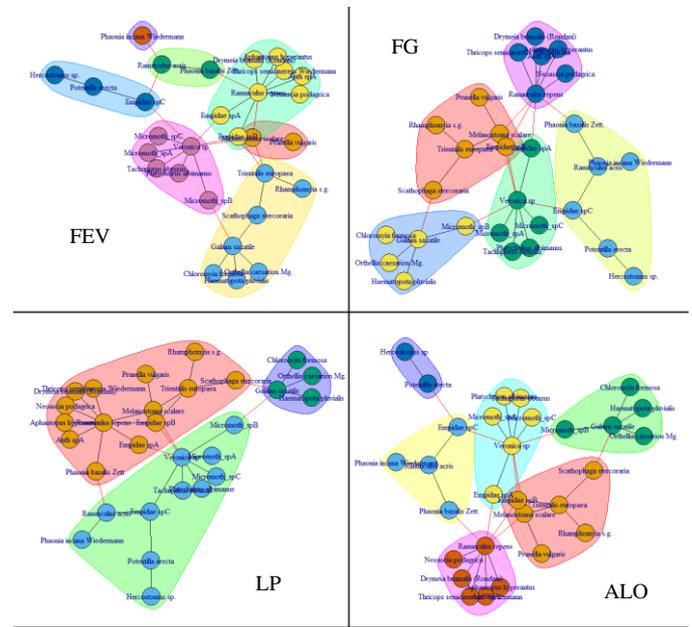
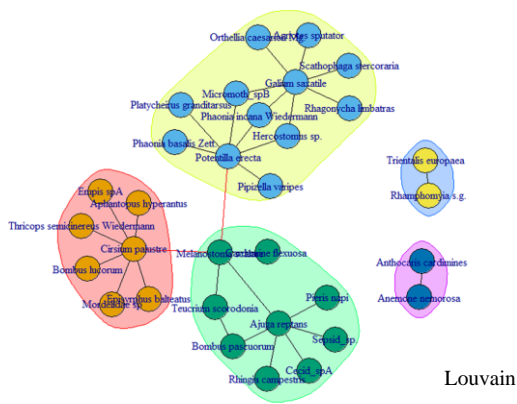
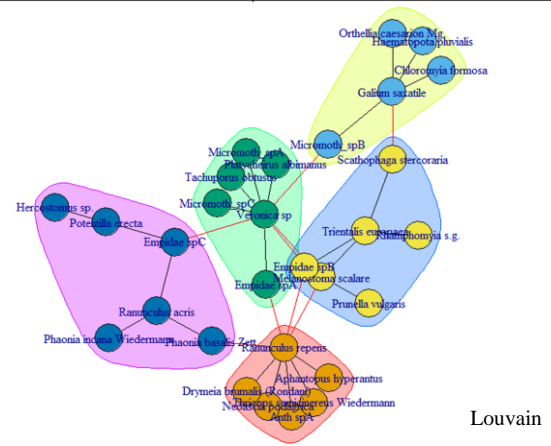

Figure 15 – Communities Detection for backfill site based on Louvain, Fast Greedy, Label Propagate, Ant Lion Optimizer and FEV algorithms.

Figure 16 - Communities Detection for the Dalhaike site based on Louvain, Fast Greedy, Label Propagate, Ant Lion Optimizer and FEV algorithms.

The provided table presents the calculation of the number of communities and the contract score for the Dalhaike site using five different methods. This evaluation allows for a comparison of the effectiveness of each method in terms of the identified communities and the corresponding contract scores.

Table 11 - Modularity score and the number of communities discovered in different methods for the Dalhaike site.

| C.D Methods | Louvain | Fast Greedy | Label Propagate | ALO | FEV |
|---|---|---|---|---|---|
| Modularity | 0.46 | 0.51 | 0.34 | 0.52 | 0.52 |
| Number of C | 7 | 6 | 3 | 5 | 5 |

The depicted figure showcases the communities achieved through various methods for the Dalhaike site. It provides a visual representation of the groupings obtained by each method. It is recognized that different algorithms yield varying values of modularity and communities. Except for the ALO and FEV algorithms, specifically, which yield identical results, all other algorithms exhibit varying outcomes.

## VII. CONCLUSION

The identification of key members or species in ecological networks is highly practical and significant. Centrality and node value indices are used to determine these important members from various perspectives. Species with high centrality, such as degree centrality and special vector, play a crucial role in communication and maintaining biodiversity within the network. If a species with high centrality is lost, it can result in significant damage to the insects or plants associated with it. On the other hand, species with high closeness centrality, if destroyed, will cause relatively less damage to the connected insects or plants. Species with first degree centrality, which have only one partner in the network, require special attention as they are highly dependent on their sole interaction for survival. Additionally, a proposed algorithm calculates the value score of each node, indicating its importance for biodiversity and the conservation of endangered species. Environmental experts are recommended to employ different methods to identify significant species within ecosystem networks. This approach aids in safeguarding environmental biodiversity and ensuring the stability of connections between species. In the context of socialization and grouping of insects and plants, different methods are used to categorize them. Sometimes, a more detailed and customized examination is required for specific groups within a collection, or when the number of members is too large to handle individually, and representative members are needed. For instance, insects are grouped based on their feeding type, while plants are grouped based on their pollination characteristics. Therefore, environmental experts are advised to utilize socialization methods for grouping insects or plants, enabling focused investigations or studies on specific categories. In this case, an example of an influential insect in biodiversity could be the X-type insects within Group 1, which are significant in terms of nutrition. The research is subject to certain limitations, particularly concerning the collected data. Despite efforts to maintain consistency across factors, inherent variability exists due to the natural context of the data. Additionally, the two-part nature of the insect-plant communication graph poses challenges in applying conventional analytical methods. Therefore, it is necessary to undertake further study, verification, method modification, or development of novel approaches to effectively analyze the graph. Furthermore, the unavailability of information regarding species' independence in sustaining life without their partners led to the utilization of approximated data in the co-extinction model analysis. The accuracy of future research can be enhanced by incorporating more precise data. Another significant limitation is the scarcity of research sources within this interdisciplinary field, impeding comprehensive exploration of the subject matter.

## VIII. FUTURE WORKS

For a more comprehensive analysis of changes, employ homogeneous data where all factors are identical, except for the variable being investigated. In the co-extinction models section, it is crucial to obtain data demonstrating the resilience of species in the absence of their partners. However, acquiring such data may be challenging, particularly for rare or specialized species that may exhibit vulnerability.